\documentstyle[12pt,epsf,epsfig]{article}
\def\beq{\begin{equation}}
\def\eeq{\end{equation}}
\def\bea{\begin{eqnarray}}
\def\eea{\end{eqnarray}}
\def\ba{\begin{array}}
\def\ea{\end{array}}
\def\gappeq{\mathrel{\rlap {\raise.5ex\hbox{$>$}}
{\lower.5ex\hbox{$\sim$}}}}
\def\permil{$\%\raise.20ex\hbox{$_0$}}
\def\lappeq{\mathrel{\rlap{\raise.5ex\hbox{$<$}}
{\lower.5ex\hbox{$\sim$}}}}
\begin{document}
\topmargin -1.0cm
\oddsidemargin -0.8cm
\evensidemargin -0.8cm
\pagestyle{empty}
\begin{flushright}
CERN-TH/96-171
\end{flushright}
\vspace*{5mm}
\begin{center}
{\Large \bf Dark Matter in Theories of Gauge-Mediated}\\
\vspace{0.5cm}
{\Large\bf Supersymmetry Breaking}\\
\vspace{2cm}
{\large S. Dimopoulos$^{\#,\ast}$,
G.F. Giudice$^\#$\footnote{On leave of absence from INFN, Sezione di
Padova,
Padua, Italy.} and A. Pomarol$^{\#}$}\\
\vspace{0.3cm}
$^\#$Theoretical Physics Division, CERN\\
CH-1211 Geneva 23, Switzerland\\
\vspace{0.3cm}
$^\ast$Physics Department, Stanford University\\ Stanford CA 94305,
USA\\
\end{center}
\vspace*{2cm}
\begin{abstract}
In gauge-mediated theories supersymmetry breaking originates in a  
strongly interacting sector and is communicated to the ordinary  
sparticles via SU(3)$\times$SU(2)$\times$U(1) carrying ``messenger''  
particles.    
Stable baryons of the strongly interacting supersymmetry breaking  
sector naturally weigh $\sim 100$ TeV and are viable cold dark matter  
candidates. They interact too weakly to be observed in dark matter  
detectors. 
The lightest messenger particle is a viable cold dark matter candidate  
under particular assumptions. It weighs less than 5 TeV, has zero spin  
and is easily observable in dark matter detectors.
\end{abstract}

\vfill
\begin{flushleft}
CERN-TH/96-171\\
July 1996
\end{flushleft}
\eject
\pagestyle{empty}
\setcounter{page}{1}
\setcounter{footnote}{0}
\pagestyle{plain}

An attractive feature of the minimal supersymmetric theory is
the existence of a stable particle, usually a neutralino, which can be
the dark matter of the universe. Recently
there
has been a resurgence of interest in theories where the breaking of
supersymmetry originates at low energies and is communicated to the
ordinary sparticles via the usual gauge forces 
 \cite{gaugea}--\cite{noi}. In these 
gauge-mediated theories the lightest sparticle is
the gravitino and all other sparticles, including the neutralino,
decay into it in a cosmologically short time. In theories where the
supersymmetry-breaking scale $\sqrt{F}$ is low, less than 100 TeV, the
gravitino mass $m_{3/2}\simeq 4$ eV $\times F/(100~{\rm TeV})^2$ is too  
small
to give a significant contribution to the present energy density of the
universe \footnote{ The gravitino can be a warm dark matter candidate
if $m_{3/2} \sim$ keV \cite{pri}, 
corresponding to $\sqrt{F}\sim 2\times 10^{3}$ TeV. 
A cold dark matter component can be  
obtained from non-thermal gravitinos produced
in decay processes, but only at the price of
unconventional choices of the relevant parameters \cite{mas}.}.
Such low values of $F$ are favored in theories with only one
mass scale;
in addition, the interpretation of the Fermilab  
$e^+ e^- \gamma \gamma+E_{T} \hspace{-.45cm}/\hspace{.45cm}$ event in the 
context of gauge-mediated  
theories requires $\sqrt{F} \lappeq 10^{3}$ TeV \cite{oop}.

In gauge-mediated supersymmetric theories there are 
two new sectors with possibly
stable particles which can act as cold dark matter candidates.

1) The secluded sector. This is the strongly interacting sector
in which supersymmetry is dynamically broken.

2) The messenger sector. This contains fields charged under  
the SU$_3 \times$SU$_2 \times $U$_1$ gauge interactions which communicate
supersymmetry breaking to the ordinary sparticles.

The secluded sector often has accidental global symmetries  
analogous to baryon number. The lightest secluded ``baryon'' $B_{\varphi}$  
is stable and a good candidate for cold dark matter, provided
that the scale of supersymmetry breaking is in  
the range of  $\sqrt{F} \sim100$ TeV already favoured by both theory and  
the Fermilab event.   
 The relic
abundance of $B_\varphi$
is determined from its annihilation cross section into   ``mesons" ({\it  
i.e.}
other strongly interacting particles not carrying the conserved quantum
number) lighter than $B_\varphi$. The $B_\varphi$ annihilation
occurs via the strong interactions and we can estimate an upper
bound on the cross section using unitarity. This implies
a bound on the $B_\varphi$ relic abundance \cite{kam}
\beq
\Omega_{B_\varphi}h^2\gappeq (m_{B_\varphi}/300~{\rm TeV})^2~.
\eeq
Therefore a strongly-interacting particle with a mass in the 100 TeV
range can be a good cold dark matter candidate.
Direct detection of the $B_\varphi$ is not
possible.  Particles in the secluded sector 
do not carry ordinary gauge quantum numbers, and therefore $B_\varphi$
can only interact with nuclear matter via loop diagrams mediated by
messenger fields. The resulting cross section is unobservable with
present techniques.

The lightest messenger field is also a possible candidate for cold dark
matter.
Indeed, if the supersymmetry-breaking sector contains only singlets
under the
SU$_3\times $SU$_2\times $U$_1$ gauge interactions and if there are no
direct
couplings between the ordinary and messenger sectors, then
the theory conserves
a global quantum number carried only by messenger fields. These
hypotheses
are fairly generic in models with natural flavour conservation.
The messenger quantum number is
typically
conserved also by the new interactions which generate the $\mu$ and
$B_\mu$
parameters of the Higgs sector \cite{noi}.
Therefore the lightest messenger is expected
to be stable and can populate the present universe as a relic of the
hot
primordial era.

The messenger sector
consists of pairs of chiral supermultiplets $\Phi+\bar\Phi$, each pair
describing
a Dirac
fermion with mass $M$ and two complex scalar  particles with mass
squared $M^2\pm F$. The cold dark matter candidate is the
lightest of these
scalars. Its gauge quantum numbers can be predicted if we
assume that messengers belong to complete GUT representations, as to
preserve
the success of gauge coupling unification. The requirement
of
gauge-coupling perturbativity up to the GUT scale restricts the
choice
of messenger representations to ${\bf 5}+{\overline {\bf 5}}$ or
${\bf 10}+{\overline {\bf 10}}$ of SU$_5$ or ${\bf 16}+{\overline
{\bf 16}}$
of SO$_{10}$. Gauge interactions split the mass spectrum of
messengers
belonging to the same GUT representation. For each irreducible
SU$_3\times$SU$_2\times$U$_1$ 
representation $a$, the mass parameters
  $M_a$ and $\sqrt{F_a}$ at the scale $Q$ are related to
the common value $M$ and $\sqrt{F}$ of the GUT multiplet
by a one-loop renormalization-group
scaling
\beq
\frac{F_a(Q)}{F}=\frac{M_a(Q)}{M}=
\prod_{i=1}^3\left[ \frac{\alpha_i (Q)}{\alpha_i(M_{GUT})}
\right]^{-2\frac{C_i}{b_i}}~.
\label{uno}
\eeq
Here $C=\frac{N^2-1}{2N}$ for the $N$-dimensional representation of
SU$_N$,
and $C=Y^2$ ($Y=Q-T_3$) for the $U_1$ factor. Also 
$b_i$ are the $\beta$-function
coefficients
$b_3=-3+n$, $b_2=1+n$, $b_1=11+5n/3$, and $n$ counts the messenger
contribution ($n=1,3,4$ for each ${\bf 5}+{\overline {\bf 5}}$,
${\bf 10}+{\overline {\bf 10}}$, and ${\bf 16}+{\overline {\bf 16}}$,
respectively). Perturbativity of $\alpha_{GUT}$ at $M_{GUT}$ implies
$n\le 4$.

Scalar particles within the same isospin multiplet are split at tree
level by the SU$_2$ D-terms.
Including these contributions, the mass eigenvalues
of the two spin-zero messenger states are:
\beq
M^2\pm \sqrt{F^2+(T_3-Q\sin^2\theta_W )^2M_Z^4\cos^22\beta}~,
\label{due}
\eeq
where $T_3$ and $Q$ are the corresponding third-component isospin and
electric charge, and $\tan\beta$ is the usual ratio of Higgs vacuum
expectation values.
As the tree-level splitting inside the SU$_2$ multiplet is proportional
to $M_Z^4$, it is important to include also one-loop corrections
proportional to $M_Z^2$. In the limit $F,~M^2-F \gg M_Z^2$, the
correction to the mass difference between the lightest electric charged 
($\varphi^+$) and
neutral ($\varphi^0$) messenger is
\beq
\delta (m_{\varphi^+}^2-m_{\varphi^0}^2)=\frac{\alpha}{4\pi}
M_Z^2\left[ 4 \ln \left(\frac{F}{M^2-F}\right)  -\ln \left(
\frac{M^2+F}{M^2-F}\right)
+\frac{2F}{M^2-F}\ln \left( \frac{2F}{M^2+F}\right) -4 \right] ~.
\label{looc}
\eeq
{}From this  we find that the neutral component 
is lighter than the charged one only if $F/M^2$ is very close to one,
where the one-loop correction in eq.~(\ref{looc}) is positive and
large, or if $F/M^2$ is very close to zero, where the tree-level
result in eq.~(\ref{due}) dominates.

{}From eqs.~(\ref{uno})--(\ref{looc}) we conclude
that in the
three
cases under consideration, 
{\it (i)} ${\bf 5}+{\overline {\bf 5}}$,
{\it (ii)}
${\bf 10}+{\overline {\bf 10}}$, {\it (iii)} ${\bf 16}+{\overline
{\bf 16}}$,
the lightest scalar
messenger is respectively: {\it (i)} the neutral or charged
component of
a weak doublet (depending on the value of $F/M^2$), 
{\it (ii)} a weak singlet with one unit of electric
charge, {\it (iii)} an SU$_3 \times$SU$_2 \times $U$_1$ singlet
\footnote{This 
conclusion can be evaded
if  messengers belonging to the same
GUT multiplet  have different masses at the unification scale. This
can
happen in non-minimal GUTs without spoiling gauge coupling
unification
\cite{dp}.}.

Case {\it
(ii)} is unacceptable, as it corresponds to charged dark matter
\cite{cha}.
It also
leads to overclosure of the universe for typical values
of the messenger masses. In case {\it (iii)}, the singlet $\varphi$
decouples when
it is still relativistic and largely overpopulate the universe:
\beq
\Omega_\varphi h^2\simeq 7\times
10^{10}\left(\frac{m_\varphi}{100~{\rm TeV}}\right)~.
\eeq
As the singlet decouples at temperatures close to $M_{GUT}$, it can
be
diluted by a period of inflation occurring below the GUT scale.
However
the next-to-lightest messenger, which is a charged isosinglet with
mass
$m$ and lifetime $\tau \sim 10^{10}~{\rm yrs}~ (100~{\rm TeV}/m)^5$
gives
rise to an unacceptable distortion of the diffuse cosmic ray
background
\cite{bck}.

Case {\it (i)} is the most promising one and we discuss it in some
detail. Let us consider values of $F/M^2$ such that the neutral  
component
of the weak doublet is lighter than the charged one.
The lightest messenger decouples from the thermal bath when it is
non-relativistic and its relic abundance is determined by its
annihilation cross section. As it is apparent from  
eqs.~(\ref{due})--(\ref{looc}),
the lightest messenger is almost degenerate in mass with
its isodoublet companion. Also, if $F\ll M$, the mass splitting 
between the lightest messenger and
the scalar or fermionic particles belonging to the same
supermultiplet
can be much smaller than the freeze-out temperature. In this case one
should consider the simultaneous co-annihilation \cite{gri} of all
quasi-degenerate species in the early universe. However, since all
these
particles have comparable annihilation cross sections, it is
perfectly
adequate to follow the cosmological fate of the lightest messenger
alone (see discussions in ref.~\cite{gri}).

The thermal average of the messenger annihilation cross section times
the collision velocity $v$, in the non-relativistic limit, is given
by
\begin{eqnarray}
\langle \sigma (\varphi \varphi^* \to Z^0Z^0)v\rangle &=&
\frac{g^4}{128\pi\cos^4\theta_W m_\varphi^2}
\left( 1-\frac{7}{2x} \right)\, ,\nonumber \\
\langle \sigma (\varphi \varphi^* \to W^+W^-)v\rangle &=&
\frac{g^4}{64\pi m_\varphi^2}
\left( 1+\frac{1-4\cos^2\theta_W-44\cos^4\theta_W}
{16\cos^4\theta_Wx} \right)\, ,\nonumber  \\
\sum_{i,j=1}^3\langle \sigma (\varphi \varphi^* \to H_i^0H_j^0)
v\rangle &=&
\frac{g^4}{1024\pi\cos^4\theta_W m_\varphi^2x}\, ,\nonumber  \\
\langle \sigma (\varphi \varphi^* \to H^+H^-)v\rangle &=&
\frac{g^4\cos^22\theta_W}{1024\pi\cos^4\theta_W m_\varphi^2x}\, ,\nonumber 
\\
\sum_{i=1}^3\langle \sigma (\varphi \varphi^* \to H_i^0Z^0)v\rangle  
&=&
\frac{g^4}{1024\pi\cos^4\theta_W m_\varphi^2x}\, ,\nonumber  \\
\sum_{i,j=1}^4\langle \sigma (\varphi \varphi^* \to  
\chi_i^0\chi_j^0)
v\rangle &=&
\frac{g^4}{32\pi\cos^4\theta_W m_\varphi^2}
\left[ \frac{r}{(1+r)^2}+\frac{1-12r-42r^2+4r^3+r^4}{8(1+r)^4x}
\right]\, ,\nonumber 
\\
\sum_{i,j=1}^2\langle \sigma (\varphi \varphi^* \to  
\chi_i^+\chi_j^-)
v\rangle &=&
\frac{g^4}{16\pi m_\varphi^2}
\left[\frac{r}{(1+r)^2} +\frac{1-26r^2+r^4}
{4(1+r)^4x}+\frac{\cos^22\theta_W}{16\cos^4\theta_Wx}\right]\, ,\nonumber  \\
\langle \sigma (\varphi \varphi^* \to f\bar{f})v\rangle &=&
\frac{g^4}{128\pi \cos^4\theta_W m_\varphi^2}
N_c\left[(Q\sin^2\theta_W
-T_3)^2+(Q\sin^2\theta_W)^2\right] \frac{1}{x}\, ,\nonumber  \\
\langle \sigma (\varphi \varphi^* \to {\tilde f}{\tilde f}^*
)v\rangle &=&
\frac{g^4}{256\pi \cos^4\theta_W m_\varphi^2} N_c (Q\sin^2\theta_W
-T_3)^2 \frac{1}{x} ~.
\end{eqnarray}
{}For the annihilation channels into neutral Higgs bosons ($H^0$),
neutralinos ($\chi^0$), and charginos ($\chi^\pm$), we have summed
over
all possible final states. Here $f$ ($\tilde f$) denote a generic
(s)quark
($N_c=3$) or
(s)lepton ($N_c=1$) with electric charge $Q$ and
third isospin component $T_3$. Finally $m_\varphi$ is the mass of the
lightest scalar messenger, $r=M^2/m_\varphi^2$,
 $x=m_\varphi /T$, and $T$
is the temperature of the annihilating particle.

The freeze-out temperature $T_f$ is defined as the temperature at
which
the annihilation rate is equal to the expansion rate, and it is given
by
($x_f=m_\varphi /T_f$)
\beq
x_f=\ln \left[\frac{0.076}{\sqrt{g_*}}\frac{M_{Pl}}{m_\varphi}
\left(A+\frac{B}{x_f}\right)
\frac{\sqrt{x_f}}{x_f-\frac{3}{2}}\right]~,
\eeq
where the total annihilation cross section has been parametrized as
\beq
\langle \sigma (\varphi \varphi^* \to {\rm anything}) v
\rangle =\frac{1}{m_\varphi^2}
\left( A+\frac{B}{x}\right) ~.
\eeq
We find that $x_f$ varies between 24 and 20, as we vary $m_\varphi$
between
1 and 100 TeV.

The messenger relic abundance in units of the critical density is
\beq
\Omega_\varphi h^2=\frac{8.5\times 10^{-5}}{\sqrt{g_*}}
\left(\frac{m_\varphi}{{\rm TeV}}\right)^2
\frac{x_f}{A+\frac{B}{2x_f}},
\label{omg}
\eeq
where $h$ is the Hubble constant in units of 100 km s$^{-1}$
Mpc$^{-1}$.
Here $g_*$ is the effective number of degrees of freedom in
thermal
equilibrium at the decoupling temperature and it is equal to 228.75,
if
we sum over the complete spectrum of the minimal supersymmetric
model.

If we require $\Omega_\varphi h^2 <1$, we obtain from eq.~(\ref{omg})
an upper bound on
$m_\varphi$ of about 5 TeV.
If the messenger sector consists of $n$ families of
${\bf 5}+{\overline {\bf 5}}$
that do not mix with each other, we have a conserved quantum number
for
each family and then $n$ stable particles.
In this case, $\Omega_\varphi h^2 <1$
leads to the constraint
\begin{equation}
\sum^n_{i=1}m^2_{\varphi_i}\lappeq (5\ {\rm  TeV})^2~.
\end{equation}
However
if there are mixing terms between the  $n$ families of messengers,
the heavy families will decay to the lightest one. The 5 TeV
upper 
 bound will only apply to the mass of the lightest scalar.

The mass difference $m_{\varphi^+}^2-m_{\varphi^0}^2$ between the
charged and neutral components of the SU$_2$ doublet is so small
that the $\varphi^+$ decay width is strongly suppressed by phase
space:
\beq
\Gamma(\varphi^+\rightarrow\varphi^0e^+\nu)=
\frac{G_F^2}{15\pi^3}\left(m_{\varphi^+}-m_{\varphi^0}
\right)^5~.
\label{gam}
\eeq
A late $\varphi^+$ decay will inject energetic particles in the  
primordial 
thermal bath, potentially destroying the successful predictions from
nucleosynthesis. We therefore require that the $\varphi^+$ lifetime
is shorter than about a second; eq.~(\ref{gam}) then implies
\beq
m_{\varphi^+}-m_{\varphi^0} \gappeq 5 ~{\rm MeV}~.
\label{split}
\eeq
This constraint singles out only two small regions of $F$ where the
lightest neutral messenger can be the dark matter. 
The first one corresponds
to $\sqrt{F}
\lappeq$ 350 GeV, where the tree-level splitting in eq.~(\ref{due})
is large enough to satisfy eq.~(\ref{split}). The second one
corresponds to $F/M^2\gappeq 0.95$, where the one-loop correction
in eq.~(\ref{looc}) is enhanced by a large logarithm. 
We recall that the stability of the messenger vacuum
requires $F/M^2<1$. In the whole
intermediate range of $F/M^2$, 
either $\varphi^+$ is lighter than $\varphi^0$,
or its lifetime is too long.

A lower bound on the messenger mass scale can be derived from the
negative
experimental searches on supersymmetric particles. It is therefore
necessary to check whether this bound is consistent with the
cosmological
constraint. For our purpose, the most relevant limits come from
sneutrino
and right-handed selectron searches at LEP1\footnote{The limit on the
right-handed selectron mass from LEP1.5 critically depends on the
nature
of the neutralino and therefore it is not of general validity.}.
Their masses can be expressed as a function of $M$ and $F$,
assuming that all $n$
messengers carry the same value of $F$ and $M$:
\begin{eqnarray}
{\tilde m}^2_{e_R}&=&\frac{10}{3}\left(
\frac{\alpha}{4\pi\cos^2\theta_W}
\right)^2 n\frac{F^2}{M^2}f(F/M^2)- M_Z^2\sin^2\theta_W \cos 2\beta\, ,
\label{mer}\\
{\tilde m}^2_{\nu_L}&=&\frac{1}{6}\left( \frac{\alpha}{4\pi}
\right)^2 \left(
\frac{9}{\sin^4\theta_W}+\frac{5}{\cos^4\theta_W}\right)
n\frac{F^2}{M^2}f(F/M^2)+ \frac{M_Z^2}{2}\cos 2\beta ~.
\label{mnl}
\end{eqnarray}
Here $f(F/M^2)$ describes the exact result of the two-loop
integration
\beq
f(x)=\left\{ \frac{(1+x)}{x^2}\left[\ln (1+x)-2 Li\left(
\frac{x}{1+x}
\right) +\frac{1}{2}Li\left( \frac{2x}{1+x}\right) \right]
+\left(
x\to -x \right) \right\}\, ,
\eeq
and it is normalized such that $f(0)=1$. The requirement that
eqs.~(\ref{mer})
and (\ref{mnl}) simultaneously satisfy
$\tilde{m}_{e_R},\tilde{m}_{\nu_L}
>M_Z/2$ implies
\beq
\frac{F}{M}\sqrt{nf(F/M^2)}>20~{\rm TeV} ~.
\label{bou}
\eeq

{}For $n\leq 4$ (which is the maximum
allowed by perturbativity of $\alpha_{GUT}$), this bound is
inconsistent with the cosmological condition $\Omega_\varphi h^2 <1$,
unless $F/M^2\gappeq 0.87$. It is interesting to notice
that the experimental limit in eq.~(\ref{bou}) selects the same region  
of
parameters allowed by the constraint on the $\varphi^+$ lifetime.
Indeed for $M\gappeq$ 10 TeV (with $n=4$) or $M\gappeq$ 20 TeV
(with $n=1$) and $F/M^2\gappeq 0.95$, all constraints can be
simultaneously satisfied and the lightest messenger is an acceptable
dark matter candidate.

The constraint in eq.~(\ref{bou}) can be relaxed if
the
messengers have different masses $M_i$  
($i=1,...,n$),
splittings
$F_i$ and small mixings to each other. 
The dark matter candidate can be the 
lightest of these messengers into which the rest can decay via their
small mixings. If the lightest messenger 
has a ratio $F/M$ which is significantly
smaller than the corresponding ratios for the other messengers, then
it  does not sizeably
contribute to sparticle masses and is not constrained by
eq.~(\ref{bou}).
Conversely, the contribution to the sparticle masses 
depend on messengers with large $F/M$ that
are unstable and, consequently, are not directly constrained from
cosmology. This scenario seems quite generic and decouples the 
relic abundance and
spectroscopic constraints, allowing for the possibility that
a messenger with $\sqrt{F}<$ 350 GeV  forms the dark matter.
Such small values of $F$ are not  
unnatural and do not require any fine tunings. Generic superpotentials  
often result in singlets with vanishing $F$-terms.

We have seen that the lightest messenger in the ${\bf 5}+{\overline  
{\bf 5}}$ 
SU$_5$ representation is a viable cold dark matter particle if
the mass parameters satisfy the requirements described above. This  
possibility
can be tested
in halo particle detection
experiments.
Because of its coupling to the $Z^0$ boson, the lightest messenger
$\varphi$
scatters off
nuclei with mass $m_N$, atomic number $Z$, and atomic mass $A$
with
non-relativistic cross section \cite{wit}
\beq
\sigma =\frac{G_F^2}{2\pi}\frac{m_\varphi^2m_N^2}{(m_\varphi +m_N
)^2}
\label{crs}
\left[ A+2(2\sin^2\theta_W -1)Z\right]^2~.
\eeq
This is four times larger than the scattering cross section of a
Dirac
neutrino with corresponding mass. Present limits \cite{dm}
from direct dark matter
searches exclude that a particle with 5 TeV mass and cross section
given by eq.~(\ref{crs}) could contribute to more than about 25 \% of a
standard galactic halo with local density 0.3 GeV/cm$^3$. Given the
uncertainty in the halo determination, it is still possible that the
lightest messenger with mass parameters as specified above
can play some role in the dynamics of our galaxy,
maybe in conjunction
with the dark baryonic matter which has been identified by the
observations
of gravitational microlensing in the Large Magellanic Cloud
\cite{mac}.
We should also recall that for large values of $m_\varphi$, the
momentum
transfer is no longer negligible, and unknown nuclear form factors
may
reduce the cross section in eq.~(\ref{crs}).

If the messenger parameters do not satisfy the constraints identified
above,
a stable messenger generally leads to relic overabundance.
The problem
can be solved by a late stage of inflation with a reheating
temperature
not much higher than the weak scale. Another possibility is to allow
the
decay of the messenger. In this case we have to introduce in the
theory
new couplings between the messengers and the ordinary matter,
 which break the
conserved messenger quantum number. If these couplings correspond to
renormalizable interactions, they are dangerous, as they presumably
introduce
flavour violations, spoiling the main motivation for considering
supersymmetry breaking at low energies. However such couplings
are likely to be
generated by Planckean physics and may then correspond to higher
dimensional operators. Here is a list of all the dimension-five  
operators which violate messenger number by one unit:
\begin{eqnarray}
\frac{1}{M_{Pl}}\int d\theta^2 \Big\{\bar 5_M10^3_F&,&5^2_M\bar 5_M \bar 
5_F\Big\}
\, ,\nonumber\\
\frac{1}{M_{Pl}}\int d\theta^4 \Big\{\bar 5^\dagger_M10^2_F&,&5^\dagger_M\bar 
5_F10_F\Big\}
+h.c.\, , 
\label{dimensionfive}
\end{eqnarray}
where $5_M$, $\bar 5_M$ and $\bar 5_F$, $10_F$ are respectively the
messenger and ordinary family SU$_5$ superfields.
The latter operators induce  messenger decays with a lifetime of   
$\sim 5\times 10^{-2}$ s.
{}Furthermore, they do not introduce flavour violations or,
 more important, proton decay. 
The reason is that all these operators violate  
messenger number by one unit; so proton decay requires two such  
operators and, consequently, it can be adequately suppressed. 
In general, operators of dimensionality $m$ cause lightest messenger  
decays with a
lifetime $\tau \sim (M_{Pl}/m_\varphi )^{2m-8}~10^{-28}$ s. 
It is clear  
that only dimension-five
operators
can allow for the messengers to decay before the time of nucleosynthesis.
Of course, both the cases of late inflation and late decay
do not allow for messengers to be the dark matter.

In conclusion, we have analysed here the different possibilities for
dark matter candidates in theories with gauge-mediated supersymmetry
breaking.

{\it (i)} The gravitino can have a significant abundance if we choose
 a rather large value of the supersymmetry-breaking scale 
$\sqrt{F} \sim 10^6$--$10^7$ GeV, which is not favored by either theory  
or the Fermilab event. The gravitino gives rise to warm dark matter
scenario and it is invisible in halo detection experiments.

{\it (ii)} The secluded sector can contain a dark matter candidate
which  feels the strong interactions responsible for supersymmetry
breaking. 
Particles up to about 300 TeV are allowed. Direct detection is  
impossible, as the
dark matter particle does not carry ordinary gauge quantum numbers.

{\it (iii)} The messenger sector can naturally have a conserved quantum
number corresponding to an accidental global symmetry. In this case 
the lightest messenger
is a stable scalar particle. It is neutral and satisfies the 
appropriate cosmological 
constraints only in  specific cases. It must belong to
the ${\bf 5}+{\overline {\bf 5}}$ 
SU$_5$ representation, be lighter than about 5 TeV, and either  
correspond
to values of $F/M^2$ very close to one ($F/M^2\gappeq 0.95$)
or to values of $F/M$
much smaller than those which correspond to other messenger fields
($\sqrt{F}<$ 350 GeV). The former option appears fine tuned whereas the  
latter is quite generic.
Direct detection experiments already constrain the contribution of such
particles to the standard local halo density to less than about 25  
{\%}.
If the lightest messenger particle does not satisfy the specific  
criteria
described above, then it leads to overclosure of the universe. The  
simplest
solution to this
problem is to introduce dimension-five Planck suppressed operators  
which violate the messenger number and allow the lightest messenger to  
decay before the time of nucleosynthesis. 

It is a pleasure to thank Michael Dine,  Gia Dvali,
Scott Thomas and Jim Wells for  
very valuable conversations.
 
\newpage

\def\ijmp#1#2#3{{\it Int. Jour. Mod. Phys. }{\bf  #1~}(19#2)~#3}
\def\pl#1#2#3{{\it Phys. Lett. }{\bf B#1~}(19#2)~#3}
\def\zp#1#2#3{{\it Z. Phys. }{\bf C#1~}(19#2)~#3}
\def\prl#1#2#3{{\it Phys. Rev. Lett. }{\bf #1~}(19#2)~#3}
\def\rmp#1#2#3{{\it Rev. Mod. Phys. }{\bf #1~}(19#2)~#3}
\def\prep#1#2#3{{\it Phys. Rep. }{\bf #1~}(19#2)~#3}
\def\pr#1#2#3{{\it Phys. Rev. }{\bf D#1~}(19#2)~#3}
\def\np#1#2#3{{\it Nucl. Phys. }{\bf B#1~}(19#2)~#3}
\def\mpl#1#2#3{{\it Mod. Phys. Lett. }{\bf #1~}(19#2)~#3}
\def\arnps#1#2#3{{\it Annu. Rev. Nucl. Part. Sci. }{\bf
#1~}(19#2)~#3}
\def\sjnp#1#2#3{{\it Sov. J. Nucl. Phys. }{\bf #1~}(19#2)~#3}
\def\jetp#1#2#3{{\it JETP Lett. }{\bf #1~}(19#2)~#3}
\def\app#1#2#3{{\it Acta Phys. Polon. }{\bf #1~}(19#2)~#3}
\def\rnc#1#2#3{{\it Riv. Nuovo Cim. }{\bf #1~}(19#2)~#3}
\def\ap#1#2#3{{\it Ann. Phys. }{\bf #1~}(19#2)~#3}
\def\ptp#1#2#3{{\it Prog. Theor. Phys. }{\bf #1~}(19#2)~#3}

\vfill\eject

\end{document}